# Ionisation Models for Nano-Scale Simulation

Hee Seo, Maria Grazia Pia, Paolo Saracco and Chan Hyeong Kim

*Abstract*–Two theory-driven models of electron ionization cross sections, the Binary-Encounter-Bethe model and the Deutsch-Märk model, have been design and implemented; they are intended to extend the simulation capabilities of the Geant4 toolkit. The resulting values, along with the cross sections included in the EEDL data library, have been compared to an extensive set of experimental data, covering more than 50 elements over the whole periodic table.

## I. INTRODUCTION

A variety of experimental applications require the capability of simulating electron interactions over a wide range - from the nano-scale to the macroscopic one: some experimental examples are the ongoing R&D (research and development) for nanotechnology-based particle detectors, plasma physics, radiation effects on semiconductor devices, biological effects of radiation etc.

General-purpose tools for electron transport are available and well established in all Monte Carlo codes based on condensed and mixed transport schemes, whereas in the lower energy end track structure codes provide simulation capabilities limited to a single, or a small number of target materials.

New developments are presented here, which intend to endow a large scale Monte Carlo system for the first time with the capability of simulating electron impact ionisation down to the scale of a few tens of electronvolts for any target element. The models are suitable for use with Geant4 [1][2].

Two theory-driven models of electron ionisation cross sections, the Binary-Encounter-Bethe [3] and the Deutsch-Märk [4] one, have been implemented; they are applicable to any target elements. The resulting values have been compared to an extensive set of experimental data, covering more than 50 elements over the whole periodic table.

## II. CROSS SECTION CALCULATION

The Binary-Encounter-Dipole (BED) [3] model was first proposed by Kim and Rudd to calculate electron impact ionisation cross sections. The Binary-Encounter-Bethe (BEB) model was elaborated as a simplification of the BED model in cases where some components of the BED cross section model would be difficult to calculate or to measure experimentally.

The BEB model involves three atomic parameters for each subshell of the target atom: the electron binding energy, the average kinetic energy and the electron occupation number of the subshell. This model does not contain any empirical or adjustable parameter.

The Deutsch-Märk (DM) model has its origin in a classical binary encounter approximation derived by Thomson [5] and its improved form of Gryzinski [6]. Its formulation involves some parameters (weighting factors), which derive from fits to experimental data; values of these parameters are reported in the original authors' publications concerning the model.

The Evaluated Electron Data Library (EEDL) [7] is exploited in the low energy electromagnetic package [8][9] of Geant4. It includes electron ionization cross sections in the energy range between 10 eV and 100 GeV; nevertheless, due to intrinsic limitations of accuracy highlighted by EEDL's authors, the use of the Geant4 low energy models based on it is recommended for incident particle energies above 250 eV. To the best of our knowledge, hardly any evidence has been documented in the literature of the accuracy of EEDL for electron energies below 1 keV; this recommendation appears to be motivated by an educated guess, rather than demonstrated by experimental validation.

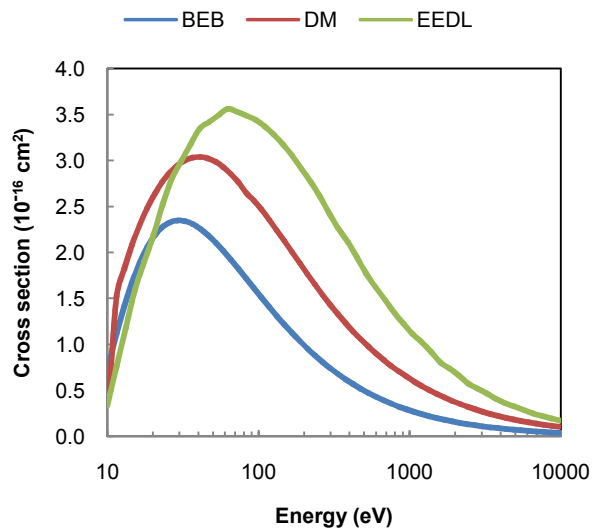

Fig. 1. Electron impact ionization cross sections for a copper target calculated by different models: Binary-Encounter-Bethe (BEB), Deutsch-Märk (DM) and the Evaluated Electron Data Library (EEDL).

The three cross section models provide different results, as shown in Fig. 1.

Manuscript received November 17, 2010. This work was partly supported by the Nuclear Research and Development Program in Korea through the Radiation Technology Development Program and the Basic Atomic Energy Research Institute (BAERI). Support also was received from the Korean Ministry of Knowledge Economy (2008-P-EP-HM-E-06-0000)/the Sunkwang Atomic Energy Safety Co., Ltd.

H. Seo and C. H. Kim are with the Department of Nuclear Engineering, Hanyang University, Seoul 133-791, Korea (e-mail: shee@hanyang.ac.kr; chkim@hanyang.ac.kr).
M. G. Pia and P. Saracco are with INFN Sezione di Genova, Via Dodecaneso 33, I-16146 Genova, Italy (telephone: +39 010 3536328, e-mail: MariaGrazia.Pia@ge.infn.it, Paolo.Saracco@ge.infn.it).

## III. SOFTWARE DEVELOPMENT

The software adopts a policy-based class design, which has also been exploited in recent developments [10][11] for photon interactions. The adoption of this design approach contributes to the ongoing investigation about the use of generic programming techniques in the physics domain of Monte Carlo simulation; feedback about its use in modeling charged particle interactions is helpful in the current R&D phase.

The policy relevant to this context is associated with a *CrossSection* function, whose arguments characterize the involved incident particle and target.

The software implementation is based on the most recent documented analytical formulations and associated parameters of the BEB and DM models, which could be retrieved in the literature at the time of writing this paper.

The formulation of both models involves some atomic parameters. Their values were taken from the same sources documented by the original authors, whenever possible; otherwise, in the cases were the original parameters were not at reach, values tabulated in the Evaluated Atomic Data Library (EADL) [12] or available from the NIST web site were used.

The implemented models allow the calculation of ionization cross sections for any element.

## IV. VERIFICATION AND VALIDATION

Verification tests were performed to check whether the cross section values calculated by the software were consistent with those calculated by the original authors of the models, which are documented in the literature.

In most cases the software implementation reproduces the original values consistently; in a few cases some discrepancies were observed, which could be tracked to different values of model parameters in the software implementation and in the original calculations. An example of these verifications is shown in Fig. 2 and Fig. 3.

As a result of the verification process, the software implementation was acknowledged to render the original cross section values with adequate precision. Further details of the verification process will be available in a dedicated paper after this conference.

The validation process involved the comparison with experimental data. A survey in the literature identified more than one hundred sets of experimental data concerning electron ionization cross sections in the low energy range below 1 keV, which are pertinent to more than 50 target elements. The quality of the experimental data is highly variable over the different samples; measurements in patent disagreement are documented in the literature.

The validation also concerned the ionization cross sections tabulated in EEDL, which is exploited by various Monte Carlo codes, including Geant4. To the best of our knowledge, this is the first time that EEDL is subject to extensive experimental benchmarks below 1 keV. Some examples of comparisons with experimental data are shown in Fig. 4-6.

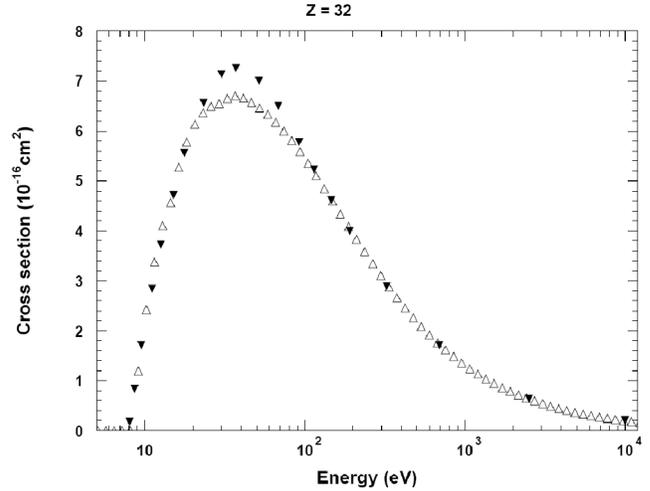

Fig. 2. An example of verification of the implementation of DM cross sections, concerning germanium (Z=32) as target: values documented by the original authors of the model (black triangles), values deriving from the software implementation (empty triangles). The small differences visible in the plot have been tracked down to the use of different weighting factors in the calculation of the cross sections. The software implementation uses the most recent values of these parameters available in the literature.

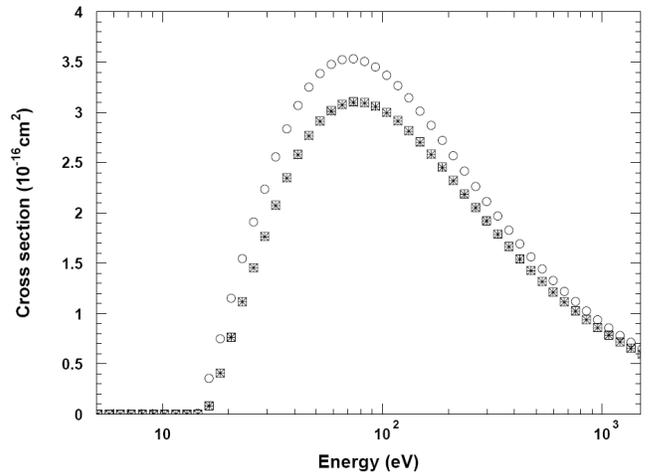

Fig. 3. An example of verification of the implementation of BEB cross sections, concerning argon (Z=18) as target: original values provided by the authors (asterisks), values deriving from the software implementation (squares) and values obtained using ionization energies as in EADL instead of those supplied by NIST (circles).

The validation test exploited rigorous statistical analysis methods [13][14] to estimate quantitatively the compatibility between the new simulation models, EEDL data and experimental data.

The validation process involved two stages: first goodness-of-fit tests to evaluate the hypothesis of compatibility with experimental data, then categorical analysis exploiting contingency tables to determine whether the various modelling options differ significantly in accuracy. Contingency tables were analyzed with the $\chi^2$ test and with Fisher's exact test.

The comparisons between the results of the implementation and experimental data were performed over selected energy ranges to verify any dependence of the modelling accuracy on the application energy. In particular, tests were performed to evaluate the degree of EEDL accuracy below 1 keV and to verify if it indeed degraded below 250 eV.

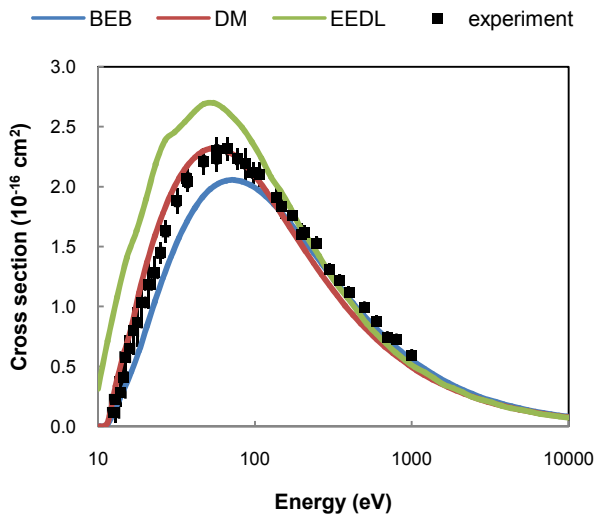

Fig. 4. Ionization cross section by electron impact on carbon (Z=6): BEB model (blue line), DM model (dark red line), EEDL (green line) and experimental data [15] (black symbols).

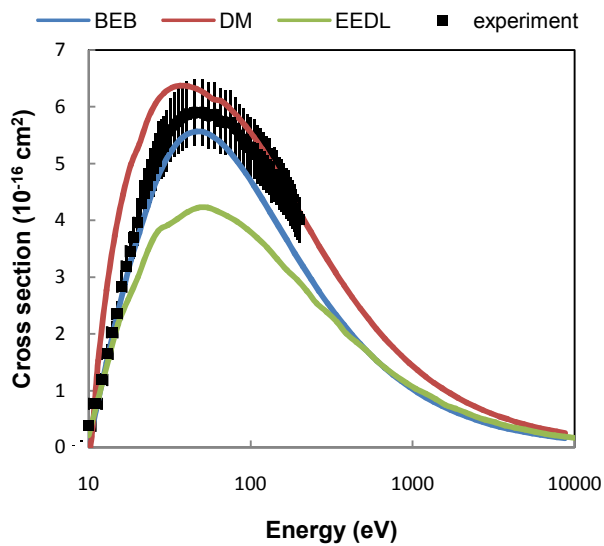

Fig. 5. Ionization cross section by electron impact on selenium (Z=34): BEB model (blue line), DM model (dark red line), EEDL (green line) and experimental data [16] (black symbols).

Further tests were performed to investigate whether the validation results could be biased by characteristics of the experimental data. The experimental references included data deriving from different types of measurements – cross sections for single or total ionization, absolute cross section measurements or relative to other data sources; tests were performed on each data category separately, and their outcome was evaluated with statistical methods to ascertain any dependence of the software accuracy on different types of experimental conditions.

According to the results of the tests, the DM model exhibits the best accuracy with respect to experimental data over the whole energy range and all the target elements subject to test. Its predictions are found to be compatible at 95% confidence level with experimental measurements for a fraction of tested target elements varying between 78% and 93%, depending on the energy range of the interacting electron.

The BEB model is comparable in accuracy to the DM model for electron energies below 100 eV at 95% confidence level.

EEDL cross sections are compatible at 95% confidence level with experimental data above 250 eV and equivalent in accuracy to the DM ones in the energy range between 250 eV and 1 keV.

The conclusions of the statistical data analysis comparing the accuracy of the various cross section options hold whatever type of data is considered: single or total ionization, absolute or relative measurements.

The results of the validation process will be documented in detail in a dedicated paper after this conference.

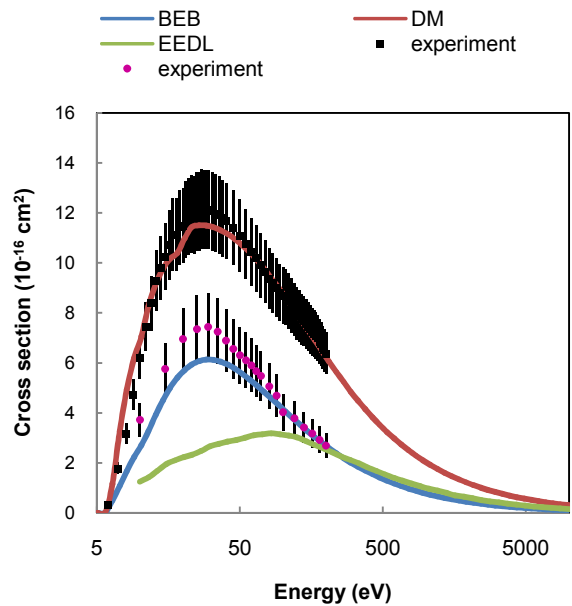

Fig. 6. Ionization cross section by electron impact on indium (Z=49): BEB model (blue line), DM model (dark red line), EEDL (green line) and experimental data [17][18] (black and purple symbols).

V. CONCLUSION

Two models for the calculation of electron impact ionization, the Binary-Encounter-Bethe and the Deutsch-Märk model, have been implemented; they are specialized for application in the low energy domain below 1 keV. The software is intended for use with Geant4; it extends Geant4 simulation capabilities in an energy range not yet covered by other general purpose Monte Carlo codes.

The software has been subject to rigorous validation with respect to a large collection of experimental measurements, concerning more than 50 target elements.

These new development open for the first time the possibility of performing microdosimetry simulations for any target elements in a general purpose Monte Carlo system.

Further developments are in progress to exploit these new cross section developments in simulation applications.

The validation project involved ionization cross sections included in EEDL as well. The accuracy of EEDL for electron energies below 1 keV has been quantitatively evaluated. These results are relevant to the use of currently available Geant4 models based on EEDL in experimental applications.

The complete set of results is documented and discussed in depth in a dedicated paper.

ACKNOWLEDGMENT

The authors express their gratitude to CERN for support to the research described in this paper.

The authors thank Sergio Bertolucci, Elisabetta Gargioni, Simone Giani, Vladimir Grichine, Berndt Grosswendt, Andreas Pfeiffer and Lina Quintieri for valuable discussions.

CERN Library's support has been essential to this study; the authors are especially grateful to Tullio Basaglia.

REFERENCES

[1] S. Agostinelli et al., "Geant4 - a simulation toolkit", *Nucl. Instrum. Meth. A,* vol. 506, no. 3, pp. 250-303, 2003.
[2] J. Allison et al., "Geant4 Developments and Applications", *IEEE Trans. Nucl. Sci.,* vol. 53, no. 1, pp. 270-278, 2006.
[3] Y. K. Kim, M. E. Rudd, "Binary-encounter-dipole model for electron impact ionization," *Phys. Rev. A,* vol. 50, pp. 3954–3967, 1994.
[4] H. Deutsch, T. D. Märk, "Calculation of absolute electron impact ionization cross-section functions for single ionization of He, Ne, Ar, Kr, Xe, N and F", *Int. J. Mass Spectrom. Ion Processes,* vol. 79, pp. R1, 1987.
[5] J. J. Thomson, "Ionization by moving electrified particles", *Philos. Mag.,* vol. 23, pp. 449-457, 1912.
[6] M. Gryzinski, "Two-Particle Collisions," Phys. Rev. A, vol. 138, pp. 305-321, 1965.
[7] S. T. Perkins et al., Tables and Graphs of Electron-Interaction Cross Sections from 10 eV to 100 GeV Derived from the LLNL Evaluated Electron Data Library (EEDL)", UCRL-50400 Vol. 31, 1991.
[8] S. Chauvie, G. Depaola, V. Ivanchenko, F. Longo, P. Nieminen and M. G. Pia, "Geant4 Low Energy Electromagnetic Physics", Proc. Computing in High Energy and Nuclear Physics, Beijing, China, 337-340, 2001.
[9] S. Chauvie et al., "Geant4 Low Energy Electromagnetic Physics", in Conf. Rec. 2004 IEEE Nucl. Sci. Symp., N33-165.
[10] M. Augelli et al., "Research in Geant4 electromagnetic physics design, and its effects on computational performance and quality assurance", 2009 IEEE Nucl. Sci. Symp. Conf. Rec., pp. 177– 180, 2009.
[11] M. G. Pia et al., "Design and performance evaluations of generic programming techniques in a R&D prototype of Geant4 physics", J. Phys. Conf. Ser., vol. 219, pp. 042019, 2010.
[12] S. T. Perkins et al., Tables and Graphs of Atomic Subshell and Relaxation Data Derived from the LLNL Evaluated Atomic Data Library (EADL), Z=1-100, UCRL-50400 Vol. 30, 1997.
[13] G. A. P. Cirrone et al., "A Goodness-of-Fit Statistical Toolkit", *IEEE Trans. Nucl. Sci.,* vol. 51, no. 5, pp. 2056-2063, 2004.
[14] B. Mascialino, A. Pfeiffer, M. G. Pia, A. Ribon, and P. Viarengo, "New developments of the Goodness-of-Fit Statistical Toolkit*", IEEE Trans. Nucl. Sci.,* vol. 53, no. 6, pp. 3834-3841, 2006.
[15] E. Brook, M. F. A. Harrison, A. C. H. Smith, "Measurements of the electron impact ionisation cross sections of He, C, O and N atoms," *J. Phys. B: Atom. Molec. Phys.,* vol. 11, no. 17, pp. 3115–3132, 1978.
[16] R. S. Freund, R. C. Wetzel, R. J. Shul, T. R. Hayes, "Cross-section measurements for electron-impact ionization of atoms", *Phys. Rev. A,* vol. 41, no. 7, pp. 3575–3595, 1990.
[17] R. J. Shul, R. C. Wetzel, R. S. Freund, "Electron impact ionization cross section of the Ga and in atoms," *Phys. Rev. A,* vol. 39, no. 11, pp. 5588–5596, 1989.
[18] L. A. Vainshtein et al., "Cross sections for ionization of gallium and indium by electrons," *Sov. Phys. JETP,* vol. 66, no. 1, pp. 36–39, 1987.